# Leveraging Large-scale Computational Database and Deep Learning for Accurate Prediction of Material Properties

Pin Chen[1], Jianwen Chen[1], Hui Yan[1], Qing Mo[1], Zexin Xu[1], Jinyu Liu[1], Wenqing Zhang[2,3], Yuedong Yang[1,*], Yutong Lu[1,*]

[1]National Supercomputer Center in Guangzhou, School of Computer Science and Engineering, Sun Yat-sen University, 132 East Circle at University City, Guangzhou 510006, China
[2]Department of Physics and Shenzhen Institute for Quantum Science & Engineering, Southern University of Science and Technology, Shenzhen, Guangdong 518055, China
[3]Guangdong Provincial Key Laboratory of Computational Science and Material Design, and Shenzhen Key Laboratory of Advanced Quantum Functional Materials and Devices, Southern University of Science and Technology, Shenzhen, Guangdong 518055, China
e-mail: luyutong@mail.sysu.edu.cn, yangyd25@mail.sysu.edu.cn

**Abstract**: Accurately predicting the physical and chemical properties of materials remains one of the most challenging tasks in material design, and one effective strategy is to construct a reliable data set and use it for training a machine learning model. In this study, we constructed a large-scale material genome database (Matgen) containing 76,463 materials collected from experimentally-observed database, and computed their bandgap properties through the Density functional theory (DFT) method with Perdew-Burke-Ernzehof (PBE) functional. We verified the computation method by comparing part of our results with those from the open Material Project (MP) and Open Quantum Materials Database (OQMD), all with PBE computations, and found that Matgen achieved the same computation accuracy based on both measured and computed bandgap properties. Based on the computed properties of our comprehensive dataset, we have developed a new graph-based deep learning model, namely CrystalNet, through our recently developed Communicative Message Passing Neural Network (CMPNN) framework. The model was shown to outperform other state-of-the-art prediction models. A further fine-tuning on 1716 experimental bandgap values (CrystalNet-TL) achieved a superior performance with mean absolute error (MAE) of 0.77 eV on independent test, which has outperformed the pure PBE (1.14~1.45 eV). Moreover, the model was proven applicable to hypothetical materials with MAE of 0.77 eV as referred by computations from HSE, a highly accurate quantum mechanics (QM) method, consist better than PBE (MAE=1.13eV). We also made material structures, computed properties by PBE, and the CrystalNet models publically available at https://matgen.nscc-gz.cn.

## Main

Experimentally observed data has always been an important resource for studying the chemical and physical properties of materials, but experiments are time-consuming and expensive. It has become popular to obtain material properties through computational methods,[1-4] and the quantum mechanics (QM) provides a solid theoretical foundation for building the structure-property relationship of materials. When a material is determined with the geometric structure in crystal, it can be computed with the electronic distribution, magnetism, optics, mechanics, and other properties. Among these, the crystal bandgap is one of the most important properties because of the wide applications (including electronic devices[5], solar cells[6], illumination[7], etc.) and the availability of abundant experimental data. As full QM computation is expensive, its approximate computation, Density Functional Theory (DFT)[8], is becoming a mature technique to compute geometric structures and physicochemical properties of experimentally-observed and hypothetical materials. Unfortunately, the involved computational costs remained high, limiting the computation on personal computers.

With the development of modern parallel packages and supercomputers, a few large-scale material databases have been recently constructed to compute properties for tens of thousands of materials using DFT computational methods, allowing scientists to screen out potential materials based on specified properties. As experimentally-observed structures are limited in material science, most of these database use hypothetical materials as an important part. For example, Material Project (MP)[9] consists of over 130,000 inorganic materials, while less than 50,000 materials are experimental materials (access time: Sep. 2020) with the Inorganic Crystal Structure Database (ICSD)[10, 11]. Automatic-FLOW for Materials Discovery (AFLOW)[12] contains over 3.5 billion materials with only about 30,000 experimentally observed materials. Open Quantum Materials Database (OQMD)[13] contains approximately 300,000 calculated structures (access time: Oct. 2020), of which about 10% of structures are from ICSD and 90% from iterations over many chemistries for several simple prototypes. Other examples are Joint Automated Repository for Various Integrated Simulations (JARVIS)[14], Novel Materials Discovery project (NOMAD)[15], MaterialGo[16] and the Materials Informatics Platform with Three-Dimensional Structures (MIP-3d)[17]. While it remains to be determined whether the hypothetical materials can be synthesized in real-world applications, it's essential to construct a comprehensive material database with experimentally observed structures.

On the other hand, the availability of such large database has also inspired scientists to adopt machine learning (ML) techniques to discover and design new materials. For molecules and crystals, graph-based neural networks have demonstrated excellent prediction performance on a broad array of properties. For example, Xie and Grossman[18] proposed a generalized crystal graph convolutional neural network

(CGCNN) for representing crystalline materials, which can accurately predict material property within DFT accuracy. Schütt et al.[19] developed a deep learning model SchNet using continuously-filtered convolutional layers for predicting molecules and materials properties. Chen et al.[20] proposed a universal graph network framework MEGNet by incorporating global state attributes. While these predictive models have been proven to efficiently design material candidates for experimental validation, the used message passing neural network (MPNN) framework[21] simply diffuses information along atoms, failing to effectively capture the message interactions between atoms and bonds. This problem was magnified in crystal because crystal structures consist of higher connectivity between atoms due to the metal and ionic bonds. Recently, our developed communicative message passing neural network (CMPNN) [22] was shown to improve the molecular embedding by strengthening the message interactions between atoms and bonds through a communicative kernel. Thus, it's promising to improve material properties prediction through the CMPNN framework.

In this study, we have constructed a comprehensive material genome database (Matgen) based on the experimental crystal materials from the ICSD (access time: Feb. 2019) and Crystallography Open Database (COD)[23, 24] (access time: Oct. 2019), totally 93,806 de-duplicated crystal structures with integral atomic position occupancy. For these collected materials, we performed a large-scale high-throughput DFT computations, leading to the comprehensive DFT computed properties of 76,707 structures in database. Based on the computational properties, we developed a directed graph-based deep learning model (CrystalNet) for predicting material properties based on the CMPNN framework[22], which outperformed state-of-the-art models by more than 29.2%. The model was further fine-tuned on 1716 experimentally measured values, and achieved an MAE of 0.77eV by an independent test on 54 data points, which is significantly lower than the ~1.14eV by PBE-based computations (P-value=8e-17) according to pairwise t-test. More importantly, our model was proven applicable to hypothetical materials with MAE of 0.77 eV as referred to computations by HSE, a highly accurate QM method, consist better than PBE (MAE=1.13eV). Hence, our publicly available database and prediction models provide a fast and accurate way for discovering and designing new materials.

# Results

**Fig. 1 | The data collection flow, structure distributions and data accuracy evaluation of Matgen database. a,** Flow chart depicting the process of data collection. **b**, Total experimental structures in Matgen, MP and OQMD databases, divided by ICSD code. **c**, Number of appearances of various chemical species in Matgen database (the logarithmic scale in the y-axis). **d**, The distribution of the material sizes with the number of atoms in the unit cell. **e**, Matrix bubble of mean absolute error (MAE) and Pearson correlation coefficient (PCC) for Matgen, MP, OQMD and experimental values. **f**, The scatter diagram between Experiment data (Exp.) and Matgen. The confidence interval is set ot 0.95.

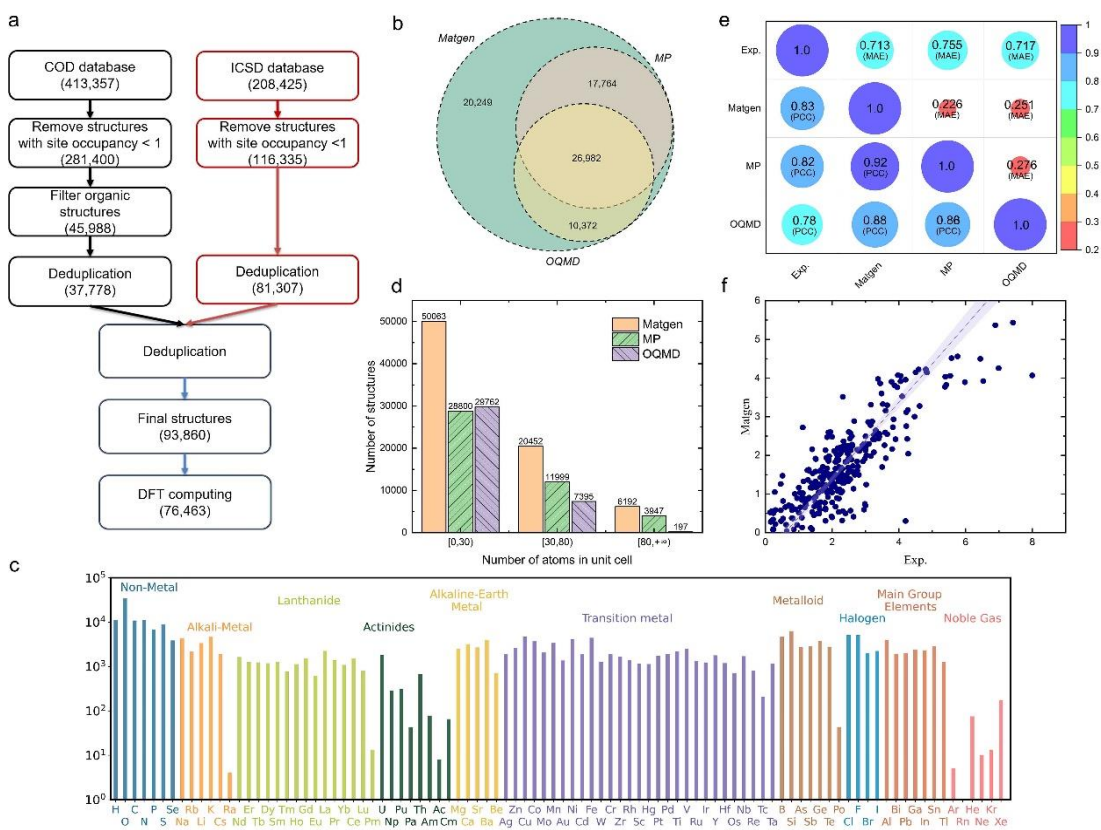

**Matgen Database Construction.** To construct a comprehensive material database with computed properties, we have culled experimental crystal structures from the Inorganic Crystal Structure Database (ICSD) and Crystallography Open Database (COD), totally 621,782 structures (Fig. 1a). After removing disordered (atom site occupancy < 1) or duplicate structures, 93,806 material structures remained. For these structures, we performed a large-scale density functional theory (DFT) computation on the Tianhe-2 supercomputer with PBE, totally costing about 100 million CPU hours on the 2.6 GHz Intel Xeon processors. After removing materials failing to converge, we finally kept computed bandgap values for 76, 463 experimental structures. The computed bandgap values (including band structure and density of state), along with Fermi energy, formation energy, magnetization, and geometry (including lattice parameters, X-ray diffraction pattern and structure factor), are publicly available in our material genome database (namely Matgen) at https://matgen.nscc-gz.cn.

As shown in Fig. 1b, Matgen is respectively 1.71 and 2.04 times greater than accessible experimental structures deposited in MP (44,746) and OQMD (37,545), currently two largest experimental material structure database. As a result, Matgen contains 20,249 (26.4%) new structures not covered by either MP or OQMD, with 26,982 common structures covered by all three databases. Matgen shows a higher diversity of chemical species by containing 94 elements (Fig. 1c), in comparison to 89 elements by MP and OQMD (Fig. S1a, Fig S1b): MP and OQMD didn't include atom types Cm, Rn, Ra, Po, and Am, although Matgen has a similar distribution of chemical species to MP and OQMD (Fig. S1c, Fig. S1d and Fig. S1e). In term of the material size (the atom number in a crystal cell), Matgen contains 1.73, 1.70, and 1.55 times more materials than MP in small (atoms $\leq$ 30), medium (30< $atoms$ $\leq$ 80), and big sized (atoms > 80) materials, respectively. Instead, OQMD is obviously biased to small sized ones (1.67, 2.85, and 41.67 times fewer than Matgen for three sizes), and contains only 129 big-sized materials. This bias is likely because the big sized materials are significantly more expensive in computations than small sized ones.

To validate our computed results, we compared our computed properties of bandgap with experimental values as well as PBE computing values in MP and OQMD database. As shown in Fig. 1e and Fig. 1f, our computed values show a Pearson correlation coefficient (PCC) of 0.830 and MAE of 0.713 to experimental ones (301 data points), slightly better than MP (0.820 and 0.755 for PCC and MAE, respectively). OQMD achieved significantly lower PCC value (0.780) likely because it used different "U" values (Fig. S2). The mutual comparison indicates our computed values are the closest to those by MP with MAE of 0.226 eV, lower than the one (0.251) between ours and OQMD and the one (0.276) between OQMD and MP. The trend is the same when expanding to the whole dataset of 26, 982 common materials shared by three PBE-based databases (Fig. S3).

**Fig. 2 | Machine learning model performance**. **a**, Communication information among nodes and edges in CrystalNet network. **b**, Plots of pairwise relationship between CrystalNet and experimental bandgap values. **c**, MAE performance of CrystalNet model for small (atoms $\leq$ 30), medium (30<atoms$\leq$80), and big sized (atoms >80) materials, **d**, MAE performance with different training data sizes, and the confidence interval is set to 0.95.and **e**, Comparison of bandgaps obtained from machine learning (ML) predictions, PBE-based values available in different database (PBE computed values), and Tran-Blah modified Becke-Johnson (MBJ) computed values compared with 54 experimental values, and more detailed values are shown in Table S3.

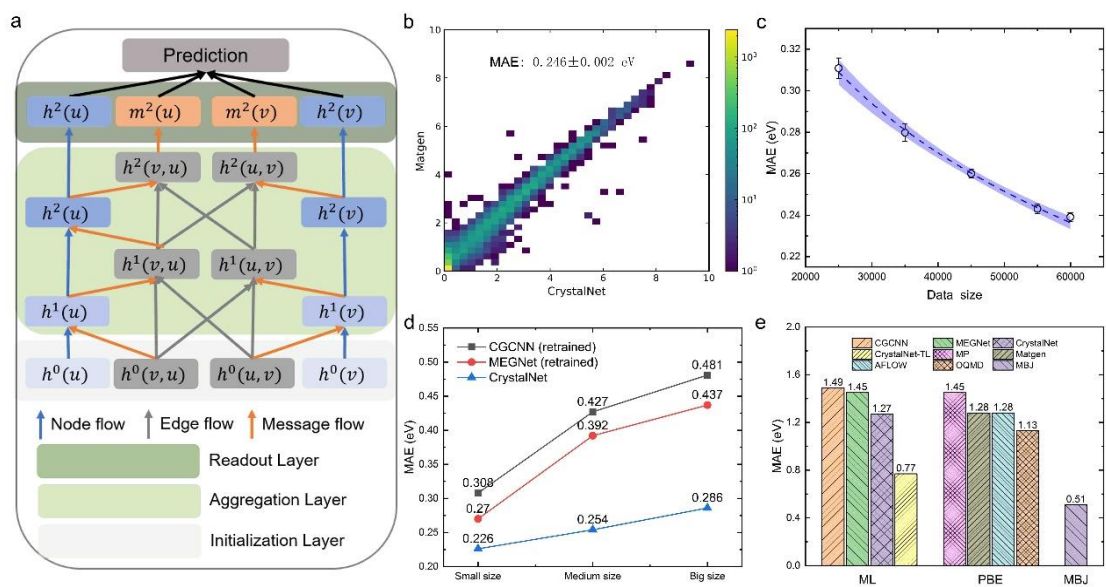

**CrystalNet Performance.** As the DFT is computationally expensive, we attempted to use computed data to train a deep learning model. We developed a directed graph-based deep learning model (CrystalNet), where the message interactions were strengthened between atoms and bonds (Fig. 2a) through a communicative message passing in between (more details available in the Methods section). As shown in Fig. 2b, the trained CrystalNet achieved an MAE of 0.246±0.002 by the independent test of 7494 data points. This is essentially the same as MAE (0.242±0.008) by the 10-fold cross validation (Fig. S4), indicating the robustness of our prediction model. By comparison, MEGNet (trained on 36,720 MP data points[20]) and CGCNN (trained on 16,485 MP data points[18]) achieved MAE of 0.33 eV and 0.388 eV, representing 34.1% and 47.7% worse performances than our model. One reason for the superior performance of our CrystalNet model is our bigger training set. As shown in Fig. 2c, when retraining our model with different number of randomly sampled training points, the average MAE increased from 0.246 to 0.301 eV with an decrease of training sample size from 60,000 to 25,000. As expected, the performance turns more fluctuated with the standard deviation (SD) of MAE is also reduced from 0.005 to 0.002.

For a direct comparison with MEGNet and CGCNN architectures, we also retrained MEGNet and CGCNN on our full-size Matgen database and found the retrained MEGNet and CGCNN could reduce the MAE from 0.33 eV and 0.388 eV to 0.318 eV and 0.346eV, respectively (Fig. S4). While the performance improvements result from the larger database and consistent DFT computing parameters in Matgen, their still lower performances than CrystalNet (0.246 eV) should be caused by the limited learning of message interactions between atoms and bonds. To indicate this, we further divided the test sets into three parts: small (atoms ≤ 30), medium (30<atoms≤80) and big (atoms >80) structures. As shown in Fig. 2d, while three models all decreased performances with an increase of the material size, CrystalNet model consistently achieved the best performances on three sizes with MAE of 0.226 eV, 0.254 eV and 0.286 eV, respectively. Relatively, MEGNet (retrained) performed worse, especially for big materials (1.19, 1.54, and 1.53 times greater MAE than CrystalNet for small, medium, and big materials). CGCNN (retrained) showed the worst prediction performance with >1.36 times greater MAE than CrystalNet.

**Fine-tuning on experimental data.** We further fine-tuned the trained CrystalNet based on 1716 experimental bandgap values, and independently tested the model on the remaining 54 data points that have been widely used by many studies[14]. It should be noted that 1716 as well as 54 experiment bandgap values are all excluded from the pre-trained data set. As shown in Fig. 2e, the fine-tuned model (CrystalNet-TL) dramatically decreased the MAE of 1.27eV by CrystalNet to 0.77 eV, which is already lower than those by PBE computation methods (1.45 eV, 1.28 eV, 1.23 eV and 1.14 eV and for MP, Matgen, AFLOW, and OQMD, respectively). Though CrystalNet-TL is still worse than MBJ and HSE (Fig. S5), these two methods are much slower than PBE, and extremely slower than our deep learning based models. Fig. S6 detailed the predicted/computed bandgap values against experimental ones for Matgen, CrystalNet, CrystalNet-TL, and HSE.

**Fig. 3 | Prediction performance on hypothetical materials. a,** The bandgap distributions of CrystalNet-TL predictions, CrystalNet predictions and PBE computed values against HSE computed bandgap values. **b,** Difference between HSE computed values and CrystalNet-TL predictions, CrystalNet predictions and PBE computed values. **c,** Different computed methods of bandgap values for $SbNO_2$ materials. **d,** Different computed methods of bandgap values for $Li_2Ti_3NiO_8$ materials.

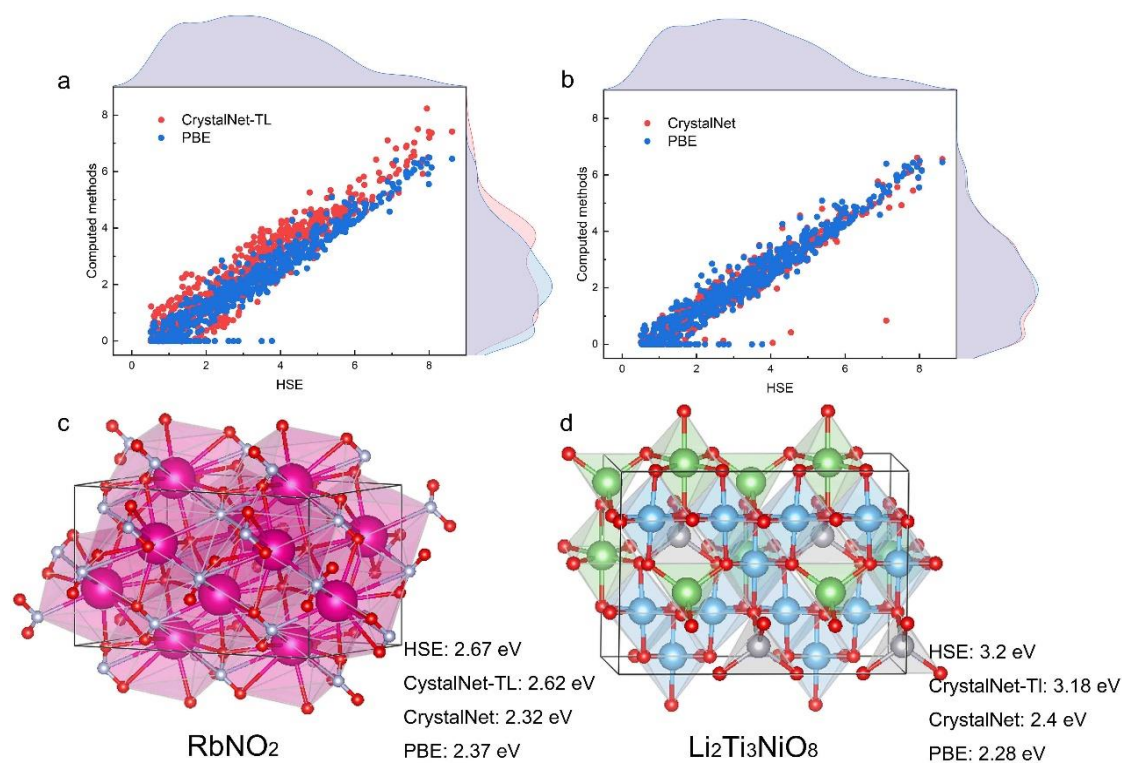

**Prediction performance on hypothetical materials.** Since experimental materials are expensive to obtain, we additionally explored the applications to hypothetical materials. We collected 602 materials from the MP database, and obtained their HSE bandgap values from the MaterialGo[16] database for reference. As shown in Fig 3a, CrystalNet predictions achieved close performance (MAE=1.17eV) to PBE calculations (MAE=1.13eV), while CrystalNet-TL significantly decreased MAE to 0.77 eV (P-value=8e-49) according to the pairwise t-test. This is likely because PBE computations underestimated the bandgap values related to HSE computing values (Fig. 3b), as also reported in previous studies[25, 26]. The fitting of PBE computation by CrystalNet caused a similar underestimation, which was alleviated by the fine-tuning in CrystalNet-TL.

As an example of the hypothetical material $RbNO_2$ (Fig. 3c), relative to the computed value (2.67 eV) by HSE, PBE and CrystalNet obtained reasonable predictions (2.37 and 2.32 eV) within DFT errors (~0.6 eV) [27], and CrystalNet-TL made a closer prediction (2.62 eV). In contrast, for the $Li_2Ti_3NiO_8$ material containing transition metals (Fig. 3d), PBE computed 2.28 eV, essentially lower than 3.2 eV by HSE. While CrystalNet followed the underestimated prediction, CrystalNet-TL corrected the prediction to 3.18 eV. Therefore, our model can be potentially used for predicting the properties of hypothetical materials to design and discover new materials.

# Discussions and Conclusion

The large set of shared and comprehensive data is one of the most important components for material innovation infrastructure to accelerate new material development. Here, we have curated the Matgen database, consisted of 76,463 experimentally-observed materials with computed properties through the Perdew-Burke-Ernzehof (PBE) method. Based on the accurate computations, we have trained a fast and accurate deep learning model through our previously developed communicative message passing neural network (CMPNN), which outperformed other state-of-the-art deep learning models. A further fine-tuning on experimentally observed properties enabled even higher accuracy than PBE, which consistently showed superior performance on hypothetical materials.

Matgen database may be an alternative to MP and OQMD database because of the following reasons. First, Matgen database is of a high quality. We have collected all structures from the experimentally-observed database (e.g. ICSD and COD) to ensure the quality of the crystal structures. The material structures in Matgen database are all collected from experimentally-observation database, ensuring the quality of the data and making the machine learning models more likely to predict material properties in real applications. Our computations were also proven consistent with MP and OQMD. Our database achieved the same or even better accuracy when compared to experimental values. Second, Matgen database contains 20,249 new experimental materials structure not covered in the MP and OQMD database because of our collection of the latest databases and supports for 94 elements. The elements of Cm, Rn, Ra, Po and Am are not found in the MP and OQMD database. Third, all the data in the Matgen database are publicly accessible on our website (https://matgen.nscc-gz.cn). Data can be obtained from API or viewed on the webpage. We provide enriched functions so that users could filter materials for their required elements. The visualization was also supported by our previously developed software, named 3DStructGen[28] with many properties shown on the webpage.

In parallel, material database can cover a limited range of materials, and the relatively slow DFT computation requires a fast and accurate substitute for material design. To address this challenge, the multi-fidelity graph networks[29] and deep transfer learning method[30] have been proposed to enhance materials property prediction, while these pioneering works showed limited performances. Here, we have developed an accurate deep learning model, CrystalNet, by fully leveraging the high quality of data in Matgen and the learning architecture in CMPNN. The model was shown to outperform previous deep learning models. We also demonstrated that the model could be further improved by fine-tuning on a small amount of experimental data, and the fine-tuned model outperformed PBE in comparison to experimental observation. This is the highlight to break through the accuracy by DFT methods through deep learning models. Notably, CrystalNet is applicable to hypothetical materials with consistent performance. Therefore, the super-fast and accurate CrystalNet provides an alternative for DFT methods to screen or design new materials in large scale.

One potential limitation of Matgen database is that the current properties are computed by PBE for a balance of accuracy and speed. Though high-fidelity DFT methods could lead to more accurate results, they are excessive computational costs. While we have proved the mode to train on low-fidelity data and fine-tune on experimental data, we might compute DFT properties for representative materials, and utilize the high-fidelity data to fine-tune models trained based on PBE computing data. Secondly, the current study was focusing on the property prediction. With the recent development of generative models like generative adversarial networks (GAN)[31, 32], we might integrate this for material design in future. Lastly, our predictions on the hypothetical materials were mostly validated through HSE. Though HSE has been widely proven, direct validations by experiments will be more promising.

The package of "Matgen + CrystalNet" presented a systematic method to improve the accuracy of material property prediction. We have proven the ability to breakthrough computational accuracy through deep learning on low-fidelity data and fine-tuning on experimental data. In future, we will further expand this mode to other material properties, and integrate the prediction for material designs.

# Methods

## Matgen database

**Crystal structures collection methods.** We first collected 621,782 structures from the union of ICSD and COD as the initial input structure source. Since we only focused on the inorganic crystal structures, 45,988 entries from the organic part of the COD database were excluded. Then, 397,735 disordered structures with partial occupancy were also removed. For these structures can only be processed by virtual crystal approximation (VCA)[33], mean-field-type coherent potential approximation (CPA)[34] or use algorithms to generate a series of unique supercells[35] to compute, which cannot be used directly in the DFT computation process. Last but not the least, we deduplicated 85,268 structures using the similarity judgement algorithm of the overlap matrix method[36]. These structures have a slight difference in initial geometrical positions caused by deviation of observation or experimental equipment, which will be likely to the same structures after geometry optimization of DFT computation. In summary, we end up with 93,806 computation-ready structures for high throughput DFT computation. Our collection methods have implemented in Matgen-toolkit package, which can be obtained from https://github.com/Matgen-project.

**Computational methods.** We performed the DFT computations with the Vienna ab initio package (VASP)[37, 38] version 5.4.4. All systems have been fully-relaxed with respect to volume as well as atomic coordinates with the generalized gradient approximation (GGA)[39] of PBE, using the pseudopotentials of the projector augmented wave (PAW) method[40, 41] at zero temperature and zero pressure. We used a unified plane wave cutoff of 520 eV for all structures to maintain the consistency and compatibility of the computed results. At the same time, this cut-off energy also satisfied the condition of 1.25 times of the maximal energy cutoff of the element plane wave basis set in all the element pseudopotentials we used. Due to the inherent approximation of the exchange-correlation energy in pure GGA may lead to poor descriptions of the electronic interaction for strongly correlated materials that contain transition metals or rare earth elements with 3d/4f orbitals. We used the GGA+U method and provide all U values of the corresponding elements in Fig. S7. We selected the Γ-centered Monkhorst-Pack (MP) scheme[42] to perform Brillouin-zone integrations. We started with k-point mesh with a dense sampling density of 2π×0.02 Å. Few of structures are difficult to converge in the computation process, we used medium-dense sampling density of 2π×0.04 Å. Besides, the blocked Davidson iteration scheme was used to solve the Kohn-Sham (KS) equations and the convergence criteria of energy and force was set to 0.1 meV and 0.001 eV/Å, respectively. We applied for about 2,000 computing nodes (24 cores per node) on the Tianhe-2 supercomputer to perform large-scale high-throughput computations. It took about 3 months to complete all the computations, and cost about 100 million CPU hours. The code of workflow for obtaining computed properties in Matgen database is available at https://github.com/Matgen-project.

## CrystalNet Method

**Crystal representation.** The key step of utilizing the graph neural network (GCN) to predict the properties is to obtain a capable chemical representation for crystals. In this field, we can get useful solutions from the chemical representation for drugs and proteins in previous researches.[43, 44] On the one hand, every crystal is stacked from many unit cells, and the unit cell is arranged by atoms with a specific packing method, which determine a crystal structure uniquely. Therefore, we can use the graph structure to represent the unit cell instead of the primary cell, with the atoms act as the nodes and the metal bonds as well as ionic bonds act as the edges. On the other hand, due to the periodicity of unit cells, the crystal graph is different from the molecule graph with fewer nodes but more edges, since it allows multiple edges between the same pair of nodes. In our CrystalNet, each node $u$ is represented by an initial feature vector $x(u)$ that collected from the atom fingerprint[18], each edge $(u,v)_k$ is also represented by a raw feature vector $x((u,v)_k)$, corresponding to the $k$th bond connecting atom $u$ and $v$. Note that the metal bonds and the ionic bonds are depended on the distance and the electronegativity between two atoms, we followed the previous work[18, 20] that expanded the distance with the Gaussian basis $\exp(-(r-r_0)^2/\sigma^2)$ centered at 100 points linearly placed between 0 and 5 and $\sigma = 0.5$. All mathematics symbols that be used in the paper can be found in Table S1.

**Deep learning framework.** Our CrystalNet was built from CMPNN[22] and specifically optimized for the crystal graph. It maintained with the edge hidden state $h((u,v)_k)$, node hidden state $h(v)$, message $m((u,v)_k)$ and $m(v)$. While CGCNN only dealt with the node hidden state $h(v)$ and edge hidden state $h((u,v)_k)$ with a simple concatenation. MEGNet only operated on the node $h(v)$ and message $m(v)$. An overall comparison among three frameworks are depicted in Fig. S8. In our network, the bond (exactly the adjacent atoms in a crystal cell) and atom embedding are updated during the training process to mimic the complex physical and chemical interactions between atoms.

The input of the CrystalNet is the directed crystal graph $G = (V, E)$, with each node was represented by concatenating all element fingerprints $x(v), \forall v \in V$ and each edge was decorated by the distance expansion vector $x((u, v)_k), \forall (u, v)_k \in E$. All node and edge raw features were act as the initial hidden embedding $h^0(v)$ and $h^0((u, v)_k)$, respectively. Then the node and edge attributes were propagated around the crystal graph during the iteration step. The output of the CrystalNet is a graph-wise vector representation $z$, which will be used as the input of multi-layer perception (MLP) for the final property prediction.

Given the iteration depth $L$, the CrystalNet began to execute aggregation step and update step alternately for $L$ times. Specifically, these two main steps consisted of four operations: i) Node aggregation, ii) Node update, iii) Edge aggregation and iv) Edge update. During the $l$-th iteration, firstly, each node $v \in V$ aggregates information from their incoming edges $\{h^{l-1}((u, v)_k), \forall u \in N(v)\}$ in $G$, creating a temporary node message $m^l(v)$ instead of using representation from its neighboring nodes, which could be written as below:

$$m^l(v) = \sum_{u \in N(v)} \sum_k Aggregate\left(h^{l-1}((u, v)_k), \forall u \in N(v)\right).$$

Note that the incoming edge information made difference between undirected and directed graph-based message passing. Besides, the aggregate function acted as a general framework of how to fuse the raw information and here we designed the multiply of sum operation and max operation as the aggregation function to maximize the potential ability of message passing, inspired by the reference[45]. Secondly, the CrystalNet utilized the node message vector to update the node's current hidden state $h^{l-1}(v)$ and generated a new hidden state $h^l(v)$ through a communicate function:

$$h^l(v) = Communicate\left(h^{l-1}(v), m^l(v)\right).$$

The new hidden state can be thought of as a message transfer station that received the incoming messages and sent an integrated one to the next station. After many experiments, we found that the add function could get the best performance.

During the edge aggregation step of CrystalNet, we generate the edge message vector $m^l((u, v)_k)$ by subtracting its inverse edge information from the $h^l(v)$, which could be shown as below:

$$m^l((u, v)_k) = h^l(v) - h^{l-1}((v, u)_k).$$

The intuition behind this minus operation is straight-forward: we have computed a high-level neighboring edges information in $h^l(v)$, which came from the previous iteration. And we tried to avoid unnecessary loops in the message passing trajectory. This step enabled the message passing from the source node to the directed edge. For the updating of edge hidden states, the edge message vector $m^l((u, v)_k)$ was feed into a fully connected layer and added with the initial $h^0((u, v)_k)$ as a skip connection following[46], and was then transformed by a rectified linear unit (ReLU) to be used at the next iteration. The formula of this step was:

$$h^l((u,v)_k) = ReLU(h^0((u,v)_k) + W \cdot m^l((u,v)_k)).$$

After iterating $L$ steps, one more round of interaction was employed to interact the enriched node and edge messages. Here, the message from incoming edges, current node hidden states, and the node's initial states were gathered to obtain the final node representation $h(v)$ of the crystal through a communicative function.

Finally, a readout operator was applied to get a fixed feature vector for the crystal. We simplified the one in MPNN as:

$$z = \sum_{v \in V} GRU(h(v)),$$

where $h(v)$ was the set of node representations in the crystal graph $G$. GRU is the Gated Recurrent Unit that introduced in this reference[47]. The final representation vector $z$ could be utilized as the input of downstream property prediction task by using such a single MLP or other tools.

**Experimental details.** To test and verify the effectiveness of the construction of Matgen database and enable head-to-head comparison of the CrystalNet model to existing crystal representation methods. We compared and analyzed all methods on our constructed dataset. Besides, for those crystals with experimental values, we made a transfer learning on those properties, to show that the fine-tuned process on the pre-trained model could make a closer and more accurate prediction. All datasets that be used in the paper can be found in Table S2.

We evaluated each method based on Matgen database. Here, we randomly selected 10% crystals as the test dataset, and the remainder was divided into 10 parts for 10-fold cross-validation. All experiments were replicated three times with different random seeds. Finally, we reported the mean and standard deviation of MAE on the cross-validation, and the final prediction performance on the independent test, respectively. The models were trained on NVIDIA Tesla v100 GPUs. On averages, it took nearly 4 minutes per epoch for each model, respectively. Most experiments reach convergence within 150 epochs. The embedding dimension was 512, the convolutional layers and the final multilayer perceptron (MLP) layers were set to 2. Besides, we used the mean square error (MSE) function as the training loss function, which could be written as below:

$$MSE = \frac{1}{n} \sum_{i=1}^{n} (y_i - \widehat{y_i})^2,$$

where the $y_i$ represented the true property value and $\widehat{y_i}$ meant the model's prediction. When coming to the evaluation process, the mean absolute error (MAE) function was calculated to measure the difference between the prediction and the label, the formula of this step was:

$$MAE = \frac{1}{n} \sum_{i=1}^{n} |y_i - \widehat{y_i}|.$$

In the transfer learning step, we firstly selected the bandgap model with the lowest error on the validation dataset, then fine-tuned the pretrained model on the 1716 experimental bandgap data points. The 1716 experimental as well as the 54 experimental data points were all excluded in our training data. We used a pre-trained CrystalNet trained on Matgen database as a starting point. The parameters between the initial layer and the last hidden layer were frozen, while parameters for the last layer were retrained. The learning rate was 1e-5 and ran 40 epochs. We used 10-fold cross-validation to train the model and then made predictions for 54 independent data points.

**Data availability**

Matgen database and prediction models are available at https://matgen.nscc-gz.cn.

**Code availability**

Our collection methods have implemented in Matgen-toolkit package, which can be obtained from https://github.com/Matgen-project/Matgen-toolkit. The high throughput computing workflow for DFT high throughput computing is available at https://github.com/Matgen-project/DFTflow. The API for obtaining computed properties in Matgen database is available at https://github.com/Matgen-project/matgen_rester. The CrystalNet model is available at https://github.com/Matgen-project/CrystalNet.

**References.**

**Acknowledgements**
This project was supported in part by Guangdong Province Key Area R&D Program (2019B010940001).
**Author contributions**
Y.L. and Y.Y. coordinated and managed this project. P.C. conceived the idea and collected the structures. J.C. implemented the deep learning model and conducted the experiments with the help of P.C.. W.Z. developed the DFT workflow method. P.C. implemented the DFT workflow code and computed the data. H.Y., Q.M., Z.X. and J.L. developed the website for displaying the computed data. P.C. and J.C wrote the manuscript with the help of Y.L. Y.Y. and W.Z.. All authors read and approved the final manuscript.
**Competing interests**
The authors declare no competing interests.
**Additional information**
Extended data is available for this paper at https://matgen.nscc-gz.cn.
Supplementary information
Correspondence and requests for materials should be addressed to Y.L. and Y.Y..